\documentclass[aps,twocolumn,superscriptaddress,pra]{revtex4}
\usepackage{bm}
\usepackage{graphicx}
\usepackage{amsbsy}
\usepackage{amsmath}
\usepackage{amsfonts}

\newcommand{\bra}[1]{\langle{#1}|}
\newcommand{\ket}[1]{|{#1}\rangle}

\newcommand{\per}{\text{per}}
\newcommand{\braket}[2]{\langle{#1}|{#2}\rangle}
\newcommand{\ketbra}[2]{|{#1}\rangle\langle{#2}|}
\newcommand{\E}{\mathcal{E}}
\newcommand{\eeqref}[1]{Eq.~(\ref{#1})}
\newcommand{\M}{N_1}

\begin{document}
\title{Efficiency limits for linear optical processing of single photons and
single-rail qubits}

\author{Dominic W. Berry}
\affiliation{School of Physical Sciences, The University of Queensland,
Queensland 4072, Australia}

\author{A. I. Lvovsky}
\affiliation{Institute for Quantum Information Science, University of Calgary,
Alberta T2N 1N4, Canada}

\author{Barry C. Sanders}
\affiliation{Institute for Quantum Information Science, University of Calgary,
Alberta T2N 1N4, Canada}

\begin{abstract}
We analyze the problem of increasing the efficiency of single-photon sources or
single-rail photonic qubits via linear optical processing and destructive
conditional measurements. In contrast to previous work we allow for the use of
coherent states and do not limit to photon-counting measurements. We conjecture
that it is not possible to increase the efficiency, prove this conjecture for
several important special cases, and provide extensive numerical results for the
general case.
\end{abstract}

\date{\today}

\maketitle

\section{Introduction}
The single-photon state of light is one of the primary resources in quantum
information technology. It is indispensable in linear optical quantum computing
\cite{LOQC1,LOQC2} and essential for many protocols of quantum communication.
However, existing single-photon sources \cite{photonsources} are far from
perfect. Whereas most ensure that the optical output contains negligible
multi-photon terms, there is always a significant probability that the desired
single photon itself is not emitted into the desired optical mode or is lost at
a later stage. As a result, the optical state generated by typical single-photon
sources can be described as an incoherent mixture of the single-photon and
vacuum states, namely
\begin{equation}\label{rho1}
\hat\rho=p\ketbra{1}{1}+(1-p)\ketbra{0}{0}.
\end{equation}
We call this state an \emph{inefficient single photon} with efficiency~$p$. The
efficiency of most existing photon sources is much lower than that desirable in
many quantum-information processing protocols.

Theoretically, the efficiency of state~(\ref{rho1}) can be improved by nonlinear
optical means, such as quantum nondemolition measurements in the photon-number
basis \cite{QNDNL1,QNDNL2}. However, this approach is not practical because
materials that combine the required nonlinearity with low optical losses are not
available at present. Therefore, it would be desirable to improve the efficiency
by means of \emph{linear} optical processing (i.e.\ interferometry) and
\emph{destructive} conditional measurements.

The possibility of such an improvement was investigated by Berry and co-workers
\cite{berry1,berry2}, who considered a general interferometric circuit with $N$
inefficient single photons as inputs. All output modes except one were subject
to photon number measurements, and the quantum state of the remaining mode,
conditioned on a particular result of this measurement, was analyzed. It was
found that the single-photon component could be increased, but at the expense of
introducing multiphoton components. It was also shown that no efficiency
improvement can be achieved if the total number of photons detected equals $0$,
$1$, or $N-1$.

A closely related problem is that of the processing of single-rail photonic
qubits. A single-rail qubit (SRQ) is a qubit with basis comprising the vacuum
state and single-photon state of one optical mode. An inefficient single photon
is a special case of a SRQ. Berry, Lvovsky, and Sanders \cite{berry} generalized
the notion of photon efficiency to SRQs, and showed that a single-rail qubit can
be converted, by means of linear optical processing and conditional
measurements, into a qubit of a different value (including the inefficient
single photon) with arbitrarily low loss of generalized efficiency. Therefore,
the problem of increasing the efficiency of single-photon sources is largely
equivalent to the problem of increasing the efficiency of SRQs.

There are a number of alternative approaches to converting a SRQ to an
inefficient single photon. A range of examples are presented in Appendix
\ref{sec:ex}. A notable feature of these schemes is that they use resources such
as coherent states and general projective measurements for the processing; these
resources were not considered in earlier work on processing single-photon
sources \cite{berry1,berry2}. Here we examine general processing including these
resources, and show that all the results found for more restricted processing
also hold when we allow coherent states and general measurements.

Another feature of the schemes presented in the Appendix is that they involve
processing of up to two sources. In previous work \cite{berry} it was only shown
that the SRQ efficiency could not be increased for processing of one SRQ. Here
we show that the same result holds for processing of up to two SRQs. In general,
we conjecture that it is not possible to improve the efficiency of SRQs or
single photons by means of linear optics and destructive measurements (without
introducing a multiphoton component). Although we do not prove this conjecture,
we do show its validity for a variety of cases, and significantly restrict the
class of apparatuses in which efficiency improvement might be achieved. We
furthermore provide extensive numerical results that support this conjecture in
the general case.

We begin by defining the general problem, and give a list of simplifications
that can be made (Sec.~\ref{sec:sim}). Then we present the general conjecture
that it is impossible to increase the efficiency, even allowing for the use of
coherent states and general measurements, in Sec.~\ref{sec:con}. In
Sec.~\ref{sec:lim} we prove the conjecture for some important special cases,
which are analogous to those studied in the case where measurements are limited
to photon counting \cite{berry1,berry2}. Numerical results in support of the
conjecture are presented in Sec.~\ref{sec:num}. In Sec.~\ref{sec:qu} we utilize
the results of previous work \cite{berry} to generalize the results to
single-rail optical qubits. Conclusions are given in Sec.~\ref{sec:cnc}.

\section{General processing and simplifications}
\label{sec:sim}
\subsection{The formulation of the problem}
\begin{figure}
\centerline{\includegraphics[width=4.8cm]{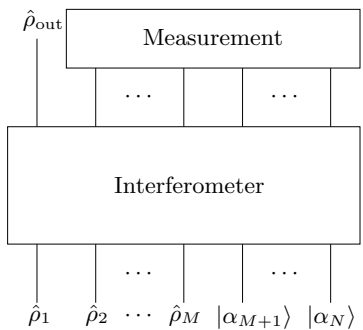}}
\caption{\label{fig:gen}General processing of multiple SRQs or single-photon
sources.}
\end{figure}

Fig.~\ref{fig:gen} displays a general scheme for processing modes by means of
linear optics and destructive measurements. In the general case the input states
$\hat\rho_1,\ldots,\hat\rho_{\M}$ are single-rail qubits. We address this case
in Sec.~\ref{sec:qu}; here we restrict these inputs to be inefficient
single-photon sources. We consider a more general form of processing than that
in previous work \cite{berry1,berry2}, and allow coherent state inputs and
general measurements on the outputs. We have $\M$ single-photon sources with
efficiencies $p_i$, and $N-\M$ coherent states $\ket{\alpha_{\M+1}},\ldots,
\ket{\alpha_N}$ (the vacuum state is also permitted by taking $\alpha_j=0$). All
inputs are processed by a linear optical interferometer, and all output channels
except channel 1 are subjected to a general destructive measurement. Conditioned
on a particular result of this measurement, we analyze the quantum state
$\hat\rho_{\rm out}$ of the remaining mode.

We restrict our treatment to schemes in which the output state has no
multiphoton ($n\ge 2$) components.  For additional generality, in the case where
coherent states are used, we allow the possibility that the multiphoton
components are small but nonzero, and take a limit where they approach zero.
This is to take account of schemes such as Scheme 2 in the Appendix, which
requires such a limit.

The goal is to either find a scheme in which the output state has an efficiency
that is higher than the highest of the input single-photon efficiencies ($p_{\rm
max}$) under this restriction on the multiphoton components, or to prove that
such a scheme does not exist. In this section, we show that \emph{if} a scheme
able to enhance the photon efficiency does exist, it can be simplified in a
number of ways without losing this property. In this way, we restrict the class
of schemes in which the desired effect should be sought. It is important that
these simplifications are performed in the order listed because, in some steps,
we assume that previous simplifications have already been made while subsequent
ones have not.

We exclude feedback or feedforward because their effect can be reproduced in a
conditional measurement setting, which is accommodated in our formalism. Because
we permit vacuum inputs and generalized measurements, we can also assume,
without loss of generality, that the interferometer is lossless and has an equal
number of input and output channels.

For this general processing the output state can be a single-rail qubit (with
coherence between the vacuum and single-photon components), rather than just an
inefficient single photon. Here we simply characterize the efficiency of the
output by the single-photon probability, rather than the SRQ generalized
efficiency \cite{berry}, which is treated later in Sec.~\ref{sec:qu}.

Next we define our notation. The input and output states of the $N$
interferometer channels are called $\hat \rho_{\rm in}$ and $\hat \rho_{\rm
trans}$, respectively. The interferometer itself is characterized by a unitary
operator $\hat U$ (so that $\hat \rho_{\rm trans}=\hat U \hat \rho_{\rm in} \hat
U^\dag$). In the Heisenberg representation, we associate the interferometer with
unitary matrix ${\bf \Lambda}$ (throughout the paper, we use boldface to
indicate vectors and matrices) such that the amplitude operators of the input
and output modes are transformed according to
\begin{equation}\label{intmap}
\hat{a}_{{\rm in},i}^\dagger \mapsto \sum_k\Lambda_{ki} \hat{a}_{{\rm out},k}^\dagger.
\end{equation}
We describe the measurement on the interferometer output modes $2,\ldots,N$ by
some positive operator-valued measure, and $\hat\rho_{\rm out}$ is conditioned
on its element $\hat M$, so that
\begin{equation}\label{rhoouttr}
\hat\rho_{\rm out}= K {\rm Tr}_{2\cdots N}(\hat M\hat\rho_{\rm trans}),
\end{equation}
where $K$ is the normalization constant that accounts for a loss of
normalization in a partial measurement of a multipartite state. The probability
that the output state contains $n_1$ photons is denoted by $c_{n_1}$ (i.e.\
$\bra{n_1}\hat\rho_{\rm out}\ket{n_1}=c_{n_1}$). We thus wish to achieve the
condition $c_1>p_{\rm max}$ and, at the same time, $\sum_{n_1>2} c_{n_1}\to 0$.

\subsection{Coherent state inputs}
\label{subsec:coh} Here we show that, without loss of generality, all the
coherent state inputs may be chosen to be vacuum states. The state with the
coherent state inputs replaced with vacuum is denoted $\hat\rho_0$. The original
input state $\hat \rho_{\rm in}$ is obtained from $\hat\rho_0$ by applying
displacement operators
\begin{equation}\label{displace}
\hat D_i(\alpha_i)=\exp(\alpha_i\hat a_i^\dag-\alpha_i^*\hat a_i)
\end{equation}
to each input $i$ ($\M< i \le N$). The input state may then be expressed as
\begin{equation}
\hat \rho_{\rm in} =\left[ \prod_{i=\M+1}^N\hat D_i(\alpha_i)\right]
 \hat\rho_0 \left[ \prod_{i=\M+1}^N\hat D_i(-\alpha_i)\right].
\end{equation}
The interferometer transforms each $\hat a_i$ according to \eeqref{intmap} and
produces the state
\begin{equation}
\hat \rho_{\rm trans} =\left[ \prod_{k=1}^N\hat D_k(\alpha'_k)\right]
 \hat U \hat\rho_0 \hat U^\dag \left[ \prod_{k=1}^N\hat D_k(-\alpha'_k)\right].
\end{equation}
where, in accordance with \eeqref{displace}, $\alpha'_k=\sum_{i=\M+1}^N \alpha_i
\Lambda_{ki}$.

Projection onto state $\ket{\chi}$ for modes 2 to $N$ yields the output state
that can be written as
\begin{equation}
\hat\rho_{\rm out}\propto \hat D_1(\alpha'_1)\left[ {\rm Tr}_{2\ldots N} (\hat
M' \hat U \hat\rho_0 \hat U^\dag) \right] \hat D_1(-\alpha'_1),
\end{equation}
where
\begin{equation}
\hat M'=\left[ \prod_{k>1}\hat D_k(-\alpha'_k)\right]\hat M \left[
\prod_{k>1}\hat D_k(\alpha'_k)\right].
\end{equation}
A displacement operator acting on a state with a finite photon number will
necessarily yield nonzero coefficients for arbitrarily large photon number. If
we require that the multiphoton components in the output are negligible, then we
must take the limit $\alpha'_1\to 0$. Our original scheme with the input state
$\hat \rho_{\rm in}$ and conditioning on measurement result corresponding to the
application of~$\hat M$ is then equivalent to a scheme with the same
interferometer, all coherent states in the input replaced by vacuum, and
measurement corresponding to~$\hat M'$.

The motivation for including the possibility of the multiphoton components
approaching zero in a limit in the above discussion is that there are cases
where this is useful when coherent states are permitted (see the Appendix for an
example). There does not appear to be any reason for allowing this possibility
in the absence of coherent state inputs, so in the remainder of this paper we
restrict the multiphoton components to be strictly zero.

Note that the above derivation does not rely on the particular form of the
states on which the displacements act. In particular, we could have included
displacements on the single-photon sources; the above derivation shows that such
displacements would not increase the power of the processing. One could
therefore define an efficiency for displaced single-photon sources (or displaced
SRQs) that is equal to the efficiency of the state without the displacement.

\subsection{Unequal efficiencies and vacuum inputs} \label{subsec:uneq}
We now make the simplification that we need not use inputs with different
efficiencies. We obtain two slightly different results.
\renewcommand{\labelenumi}{\alph{enumi})}
\begin{enumerate}
\item The maximum output efficiency may be obtained without modifying the
interferometer and taking all inputs to either have efficiency $p_{\rm max}$ or
be the vacuum.
\item For any interferometer which achieves a certain output efficiency, we may
achieve an efficiency at least as large with a modified interferometer and all
inputs with efficiency $p_{\rm max}$.
\end{enumerate}
\renewcommand{\labelenumi}{\arabic{enumi}.}
Note the difference in these results: in the first we use the \emph{same}
interferometer, and in the second we allow a modified interferometer.

To prove the first result, we note that the input state for $p_i\le p_{\rm max}$
is just a convex combination of states with all efficiencies either $p_{\rm
max}$ or 0 (vacuum). Thus the output state is again a convex combination, though
with possibly different weightings. The single-photon probability must be
maximized for one of the states in this convex combination, and therefore is
maximized for all efficiencies either $p_{\rm max}$ or 0. This result holds
regardless of whether we require the multiphoton components to be zero. Clearly
if the multiphoton components are zero for the output state with $p_i\le p_{\rm
max}$, they are zero for all states in the convex combination.

For the second result, we note that we can achieve a vacuum state simply by
combining two single-photon sources at a beam splitter and conditioning on
detection of two photons in one of the beam splitter outputs. (Note that the
beam splitter must be chosen such that the probability for detection of two
photons is nonzero.) We may use the first result to show that all inputs can
have efficiency either $p_{\rm max}$ or 0, then replace all the vacuum inputs
with two sources with efficiency $p_{\rm max}$. Thus we see that any efficiency
we can achieve with vacuum inputs can also be achieved with all inputs
identical, although with a modified interferometer.

\begin{figure}
\centerline{\includegraphics[width=8.4cm]{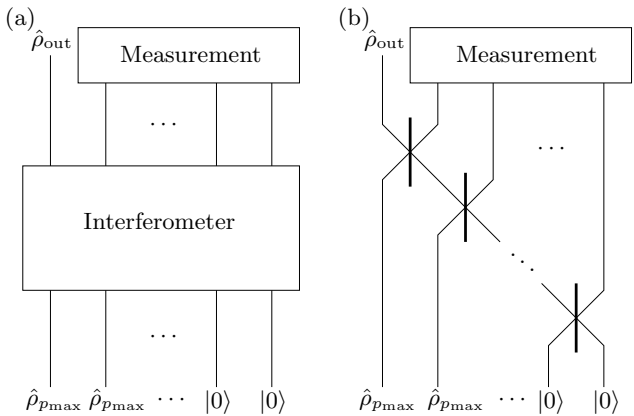}} \caption{\label{fig:sim}A
method of simplifying a scheme with multiple vacuum inputs.}
\end{figure}

We now show that, in order to obtain the second result, it is not necessary to
expand the interferometer by more than one mode. First let us consider a scheme
with all inputs with efficiency either $p_{\rm max}$ or 0, and sort all the
vacuum inputs to the right, as in Fig.\ \ref{fig:sim}(a). We may then simplify
the scheme to a line of beam splitters followed by a $U(N-1)$ interferometer on
modes 2 to $N$, then measurement on modes 2 to $N$. The $U(N-1)$ interferometer
and measurement may be combined into a single measurement, as in Fig.\
\ref{fig:sim}(b).

Now the line of beam splitters acting on the vacuum inputs simply give vacuum
outputs. Only one of these vacuum outputs is combined with non-vacuum states at
further beam splitters; the remainder remain in the vacuum state. Conditional
measurement on these modes does not affect the output state. These output modes,
and the corresponding input modes can be thus removed from the interferometer,
which simplifies the scheme to one which uses only one vacuum input.

We can replace that one vacuum input with two inputs with efficiency $p_{\rm
max}$, as discussed above. Overall the number of modes needed is just $\M+2$. If
$\M$ was equal to $N$, then there were no vacuum inputs, so we do not expand the
interferometer. Otherwise the total number of modes needed is no more than
$N+1$. Thus we can simplify the scheme to one with all identical inputs, and
need not increase the number of modes by any more than~$1$.

\subsection{General measurements}
\label{subsec:gen} The maximum output efficiency may be obtained using a
projective measurement, rather than a more general measurement operator. The
measurement on modes 2 to $N$ may be decomposed into projection operators as
$\hat M=\sum_i q_i \ket{\chi_i}\bra{\chi_i}$, so the output state will then be
\begin{equation}
\hat\rho_{\rm out} \propto \sum_i q_i \bra{\chi_i}\hat\rho_{\rm trans}\ket{\chi_i}.
\end{equation}

Thus
\begin{equation}
    \hat\rho_{\rm out}=\sum_i q'_i \hat\rho_{{\rm out},i}
\end{equation}
for some probabilities $q'_i$ with $\hat\rho_{{\rm out},i}\propto \bra{\chi_i}
\hat\rho_{\rm trans}\ket{\chi_i}$. Hence the single-photon probability for
$\hat\rho_{\rm out}$ is a convex sum over that for the $\hat\rho_{{\rm out},i}$,
and the maximum must be achieved for one of the $\hat\rho_{{\rm out},i}$.
Therefore we see that the optimal result may be obtained via projective
measurements.

\subsection{Projections over different total photon numbers}
\label{subsec:dpn} Next we show that the state $\ket\chi$ associated with the
projective measurement on modes 2 to $N$ does not need to be a superposition
over different total numbers of photons. This result holds true when there is no
coherent superposition over different total photon numbers in the input state.
We may assume this to be the case because we have shown that coherent state
inputs may be omitted. Because the interferometer does not change the total
number of photons, its output state can be written as
\begin{equation}
\hat\rho_{\rm trans} = \sum_L p_L \hat\rho_{{\rm trans},L},
\end{equation}
where $\hat\rho_{{\rm trans},L}$ denotes a state with a total of $L$ photons in
the output modes. Similarly, we can write the projection state as a
superposition
\begin{equation}
\ket\chi = \sum_D x_D \ket{\chi_D},
\end{equation}
where $\ket{\chi_D}$ is a normalized state containing terms with a total of $D$
photons. The probability that mode~1 contains $n_1$ photons is thus
\begin{align}
\label{nprob}
 c_{n_1}
        =& K\bra{n_1}\hat\rho_{\rm out}\ket{n_1}
        = K\bra{n_1}\bra\chi\hat\rho_{\rm  trans} \ket\chi\ket{n_1}  \nonumber \\
        =& K\sum_L p_L\sum_{D,D'} x_D^* x_{D'} \bra{n_1}\bra{\chi_D}
            \hat\rho_{{\rm trans},L} \ket{\chi_{D'}}\ket{n_1} .
\end{align}
We now notice that the above matrix element is nonzero only if $D=D'=L-n_1$. We
can therefore write, specifically for $n_1=1$,
\begin{equation}
c_1=K\sum_L p_L |x_{L-1}|^2 \bra{1}\bra{\chi_{L-1}}\hat\rho_{{\rm trans},L}
\ket{\chi_{L-1}}\ket{1} .
\end{equation}
This is a convex sum and must reach a maximum for $x_{L-1}$ equal to 1 for one
specific value of $L=L_0$ and 0 for others. If the projection state $\ket\chi$
is replaced by $\ket{\chi_{L_0-1}}$ (which does not contain superpositions of
different total number of photons), the value of $c_1$ will increase or remain
the same, which completes the proof.

\section{General conjecture}
\label{sec:con}
\subsection{The conjecture}
In this section, we conjecture that it is not possible to improve the single
photon efficiency provided the multiphoton component in $\hat\rho_{\rm out}$ is
zero. This conjecture may be expressed in the following way:
\begin{equation} \label{conjold}
\sum_{n_1=2}^N c_{n_1} =0 \implies c_1 \le p_{\rm max},
\end{equation}
where $p_{\rm max}$ is the highest of the input source efficiencies.

This form is somewhat inconvenient to use in numerical testing, as it has a
large number of independent efficiencies $p_i$. Therefore, in the subsequent
treatment, we invoke the simplification of Sec.~\ref{subsec:uneq} and assume
that all input channels are in the state (\ref{rho1}) with $p_i=p_{\rm max}$. In
order to formally express our conjecture, we write the circuit input as a
probabilistic mixture of pure states defined by vector ${\bm
s}=(s_1,\ldots,s_N)$, where $s_i=0$ or 1 determines whether the photon is
present in the $i^\text{th}$ input channel:
\begin{equation}
\hat \rho_{\rm in} = \sum_{\bm{s}} P_{\bm{s}} \left[ \prod_i (\hat{a}_i^\dagger)^{s_i}
\ket 0 \bra 0 \prod_i \left(\hat{a}_i\right)^{s_i} \right],
\end{equation}
where
\begin{equation}\label{Ps}
P_{\bm s}=p_{\rm max}^{\Sigma_{\bm{s}}}(1-p_{\rm max})^{N-\Sigma_{\bm{s}}}
\end{equation}
is the probability of occurrence for a particular vector ${\bm{s}}$,
$\Sigma_{\bm{s}}$ being the number of nonzero elements in ${\bm{s}}$. The
interferometer maps the operator of every mode according to \eeqref{intmap},
thus producing the state
\begin{equation}
\hat\rho_{\rm trans} = \sum_{\bm{s}} P_{\bm{s}} \prod_i \left( \sum_k
\Lambda_{ki} \hat a_k^\dagger\right)^{\!\!s_i} \!\! \ket{0}\bra{0} \prod_i
\left( \sum_k \Lambda_{ki}^* \hat a_k\right)^{\!\!s_i} \! .
\end{equation}
If modes 2 to $N$ are projected onto state $\ket\chi$, the probability that
mode~1 contains $n_1$ photons is
\begin{equation}\label{prT}
c_1=K\bra{n_1}\bra{\chi}\hat\rho_{\rm trans}\ket{\chi}\ket{n_1}=K\sum_{\bm s}
P_{\bm s} |T_{{\bm s},\chi}^{(n_1)}|^2,
\end{equation}
where we have introduced the quantity
\begin{equation} \label{Tdef}
T_{{\bf s},\chi}^{(n_1)}=\bra{n_1}\bra{\chi}\prod_i \left( \sum_k \Lambda_{ki}
\hat a_k^\dagger\right)^{s_i} \ket{0},
\end{equation}
which yields the amplitude of state $\ket{n_1}\ket\chi$ emerging in the
interferometer output provided that the input state is determined by vector
${\bm s}$.

This amplitude has the important property
\begin{equation} \label{Tprop}
T_{\bm{s},\chi}^{(n_1+1)} = \frac{1}{\sqrt{n_1+1}}\sum_{i;s_i=1}
\Lambda_{1i}T_{\bm{s}^i,\chi}^{(n_1)}
\end{equation}
with ${\bf s}^i$ a vector identical to ${\bf s}$ except the $i^\text{th}$
position at which the value~$1$ is replaced by~$0$. \eeqref{Tprop} is proven in
the next subsection. \eeqref{Tprop} implies that if $T_{\bm{s},\chi}^{(2)}=0$
for all $\bm{s}$, then $T_{\bm{s},\chi}^{(n_1)}=0$ for all $\bm{s}$ and $n_1>2$.
It then follows that if $c_{n_1} =0$ holds for $n_1=2$, it must also hold for
all $n_1>2$.

The inequality $\bra{1}\hat\rho_{\rm out}\ket{1}\le p_{\rm max}$  in
\eeqref{conjold} is equivalent to
\begin{equation}
\label{less} \sum_{\bm{s};\Sigma_{\bm{s}}=D+1} P_{\bm{s}}
|T_{\bm{s},\chi}^{(1)}|^2 \le \frac{p_{\rm max}}{1-p_{\rm max}}
\sum_{\bm{s};\Sigma_{\bm{s}}=D} P_{\bm{s}} |T_{\bm{s},\chi}^{(0)}|^2,
\end{equation}
where we have assumed, according to Sec.~\ref{subsec:dpn}, that the projection
state $\ket\chi$ has a certain total number of photons $D$. Substituting the
expression~(\ref{Ps}) for $P_{\bf s}$ into \eeqref{less}, we cancel all
probability-related factors and rewrite our conjecture in the form
\begin{equation} \label{conjT}
\sum_{\bm{s}} |T_{\bm{s},\chi}^{(2)}|^2=0 \implies \sum_{\bm{s}}
|T_{\bm{s},\chi}^{(1)}|^2 \le \sum_{\bm{s}} |T_{\bm{s},\chi}^{(0)}|^2.
\end{equation}
This form is more useful than that given in the beginning of this section,
because it does not depend on the probabilities.

\subsection{Mathematical formulation}
In this subsection, we develop a formalism that allows us to provide a
formulation of the conjecture in a pure mathematical form that does not involve
quantum amplitudes. We begin by decomposing the measurement state $\ket\chi$
into tensor products of Fock states
\begin{equation}
\ket\chi=\sum_{\bar{\bm{n}}}\chi_{\bar{\bm{n}}} \ket{\bar{\bm{n}}},
\end{equation}
where vector $\bar{\bm{n}}=(n_2,\ldots,n_N)$ determines the number of photons in
modes 2 to $N$ of the interferometer output. We can then rewrite \eeqref{Tdef}
as
\begin{equation}
T_{\bm{s},\chi}^{(n_1)}
=\sum_{\bar{\bm{n}}}\chi_{\bar{\bm{n}}}^*(\bm{n}!)^{-1/2} S_{\bm{s},\bm{n}},
\end{equation}
where $\bm{n}=(n_1,\ldots,n_N)$ is the vector $\bar{\bm{n}}$ with the addition
of the first mode, $\bm{n}!=\prod_{j=1}^N n_j!$ and
\begin{equation}
S_{\bm{s},\bm{n}}=( \bm{n}! )^{1/2} T_{\bm{s},\bar{\bm{n}}}^{(n_1)}=( \bm{n}!
)^{1/2}\bra{\bm{n}}\prod_i \left(\sum_k \Lambda_{ki} \hat
a_k^\dagger\right)^{s_i} \ket{0}.
\end{equation}
A direct calculation shows that $S_{\bm{s},\bm{n}}=\per(\bm{\Lambda}
[\bm{n},\bm{s}])$, where `per' is the permanent of a matrix, provided $\bm{s}$
and $\bm{n}$ correspond to the same total numbers of photons ($\sum_i s_i=\sum_i
n_i$) and zero otherwise. $\bm{\Lambda}[\bm{n},\bm{s}]$ is a matrix obtained
from $\bm{\Lambda}$ by repeating the $i$'th column of $\bm{\Lambda}$ $s_i$
times, and the $j$'th row $n_j$ times.

Similarly to the determinant, the permanent of a matrix can be expanded by
minors, but with all the signs taken as positive \cite{mathworld}. Therefore, if
we define $\Gamma_{\bm{s},\bar{\bm{n}}}^{(n_1)}=S_{\bm{s},\bm{n}}^*$, we can
write
\begin{equation}
\Gamma_{\bm{s},\bar{\bm{n}}}^{(n_1+1)}=\sum_{i;s_i=1}
\Lambda_{1i}\Gamma_{\bm{s},\bar{\bm{n}}}^{(n_1)},
\end{equation}
from which one immediately obtains \eeqref{Tprop}.

With the introduced notation, we are now ready to present another form of our
conjecture. By taking $\bm{\chi}$ to indicate the vector of values
$\chi_{\bar{\bm{n}}}(\bar{\bm{n}}!)^{-1/2}$, we can rewrite \eeqref{conjT} as
follows
\begin{equation}
\label{eq:con}
\bm{\chi}^\dagger {\bm{\Gamma}^{(2)}}^\dagger\bm{\Gamma}^{(2)}\bm{\chi}=0 \implies
\bm{\chi}^\dagger {\bm{\Gamma}^{(1)}}^\dagger\bm{\Gamma}^{(1)}\bm{\chi}
 \le \bm{\chi}^\dagger {\bm{\Gamma}^{(0)}}^\dagger\bm{\Gamma}^{(0)}\bm{\chi},
\end{equation}
for all vectors $\bm{\chi}$, where the product indicates summation over
indices $\bm{s}$ and $\bar{\bm{n}}$. The conjecture is therefore that
\begin{equation}
    {\bm{\Gamma}^{(1)}}^\dagger\bm{\Gamma}^{(1)}\le
    {\bm{\Gamma}^{(0)}}^\dagger\bm{\Gamma}^{(0)}
\end{equation}
on the null space of $\bm{\Gamma}^{(2)}$.

We have not yet been able to find a general proof of this conjecture. In the
following two sections, we give an analytical proof for some special cases
(which may be used as a basis for a future general proof by induction) and
report the results of numerical tests that support the conjecture.

\section{Special cases}
\label{sec:lim} The conjecture \eqref{conjold} can be proven for certain special
cases defined by the total number $D$ of photons detected in the measurement on
modes $2$ to $N$. In particular, the case $D=\M$ is trivial as it implies
$c_0=1$. The case $D=0$ involves projection onto the state
$\ket{\chi_0}=\ket{0,\ldots,0}$ and has been studied previously
\cite{berry1,berry2}. Below we prove the conjecture for $D$ equal to~$1$ and
$\M-1$. We then use these results to prove no-go theorems for $N\le 3$ or $\M\le
2$. In this section we do not assume that the inputs are all identical (that
$p_i=p_{\rm max}$ and $\M=N$), except in the proof for $D=1$.

One might think that one needs to require $D\ge N_1-1$ in order to eliminate
multiphoton components from the output. This is, however, not the case, the
simplest counterexample being a $(N=2)$-mode interferometer with
$\Lambda_{11}=\Lambda_{22}=1$, $\Lambda_{12}=\Lambda_{21}=0$ (i.e.\ direct
connection of the input and output modes). Detection of vacuum in mode 2 ($D=0$)
does not imply that mode~1 has multiphoton terms. Therefore, the proof for
$D=N_1-1$ is not sufficient to prove the conjecture.

\subsection{$D=1$}
\begin{figure}
\centerline{\includegraphics[width=8.4cm]{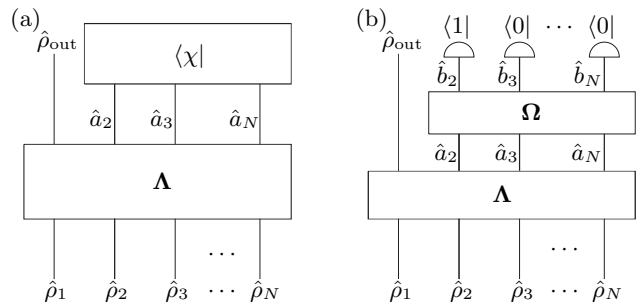}} \caption{\label{fig:D1}
Projection onto a multimode state $\ket\chi$ with a total number of photons
equal to 1 (a) can be replaced by processing with an additional interferometer
(${\bf \Omega}$) and a photon counting measurement in each output (b).}
\end{figure}

Because $\ket\chi$ contains only one photon, we can write it as a superposition
$\ket\chi=\sum_{i=2}^N \phi_i\hat{a}_i^\dagger \ket 0$. If we introduce a unitary
transformation $\{\hat a_i\} \mapsto \{\hat b_i\}$ on modes 2 to $N$ such that
$\hat b^\dagger_2=\sum_{i=2}^N \phi_i\hat{a}_i^\dagger$, we can write
\begin{equation}
\ket\chi=\hat b^\dagger_2\ket 0,
\end{equation}
so the state $\ket\chi$ corresponds to a single photon in the optical mode
defined by operator $\hat b_2$, and vacuum in the remaining modes ($\hat
b_3,\ldots,\hat b_N$). Because there exists an interferometer associated with
any unitary transformation of optical modes \cite{Zeilingerunitary}, a
projection measurement onto state $\ket\chi$ can be achieved by processing the
modes 2 to $N$ with an additional interferometer and counting photons in each
output mode (Fig.~\ref{fig:D1}).

Considering the two interferometers of Fig.~\ref{fig:D1}(b) as a single
interferometer, we find that the same output may be achieved with a modified
interferometer and photon counting. It is known that, with photon counting
measurements, $D=1$, $p_i=p_{\rm max}~\forall p_i$ and $\M=N$, it is not
possible to obtain increased efficiency \cite{berry1,berry2}. Thus we have found
that this result also holds for general projection measurements that satisfy
these conditions.

\subsection{$D=\M-1$} \label{D1}
We begin by rewriting \eeqref{prT} for the vacuum probability:
\begin{equation}
c_0 = K \sum_{\bm{s};\Sigma_{\bm{s}}=D} P_{\bm{s}} |T_{\bm{s},\chi}^{(0)}|^2.
\end{equation}
Following earlier reasoning concerning increasing the efficiency of single
photon sources by interferometry and postselection based on photon counting
\cite{berry1,berry2}, we notice that any input vector ${\bm s}$ with $D$ nonzero
elements can be obtained from $\M-D$ vectors with $D+1$ nonzero elements by
setting one of their elements to zero. We write
\begin{equation}
c_0 =  \frac {K}{\M-D}
\sum_{\substack{\bm{s}; P_{\bm{s}}\ne 0 \\
\Sigma_{\bm{s}}=D+1 }}\sum_{k;s_k=1} P_{\bm{s^k}} |T_{\bm{s^k},\chi}^{(0)}|^2.
\end{equation}
In turn this implies
\begin{equation}\label{pr0DN1}
c_0 \ge \frac{1-p_{\rm max}}{p_{\rm max}} \frac
{K}{\M-D}\sum_{\bm{s};\Sigma_{\bm{s}}=D+1}P_{\bm{s}} \sum_{k;s_k=1}
|T_{\bm{s^k},\chi}^{(0)}|^2.
\end{equation}

Now to obtain the probability for one photon, we use
\begin{equation}
c_1 = K \sum_{\bm{s};\Sigma_{\bm{s}}=D+1} P_{\bm{s}} |T_{\bm{s},\chi}^{(1)}|^2.
\end{equation}
Using \eeqref{Tprop}
we get
\begin{equation}
c_1 = K \sum_{\bm{s};\Sigma_{\bm{s}}=D+1} P_{\bm{s}} \left| \sum_{k;s_k=1}
\Lambda_{1k}T_{\bm{s^k},\chi}^{(0)} \right|^2.
\end{equation}
Using  the Cauchy-Schwarz inequality as well as the unitarity of $\bf\Lambda$
(so that $\sum_{k=1}^{N}\left| \Lambda_{1k} \right| ^2=1$), we obtain
\begin{equation}
c_1\le K\sum_{\bm{s};\Sigma_{\bm{s}}=D+1} P_{\bm{s}} \sum_{k;s_k=1}
|T_{\bm{s^k},\chi}^{(0)}|^2.
\end{equation}
Comparing this result with \eeqref{pr0DN1}, we find
\begin{equation}
\frac{c_1} {c_0} \le \frac{p_{\rm max}}{1-p_{\rm max}}(\M-D).
\end{equation}
This result reduces to previous results for measurements restricted to
projections onto tensor products of Fock states \cite{berry1,berry2}.

\subsection{$\M\le 2$ or $N\le 3$} \label{M2N3}
In the case $N\le 3$, the above no-go theorems eliminate every possibility for
an efficiency improvement. For $\M=1$, we either have $D=0$ or $D=1=\M$, so
there can be no improvement. For $\M=2$, we either have $D=0$, $D=1=\M-1$, or
$D=\M$. In each case the above no-go theorems show that no improvement is
possible. This means that no efficiency improvement is possible for $\M\le 2$,
and therefore for $N\le 2$.

In the case $N=3$, we know from Sec.~\ref{subsec:uneq} that the efficiency is
maximized either for $\M=3$ and all inputs identical, or with $\M\le 2$
(\emph{without} expanding the interferometer). We know that there is no
improvement possible for $\M\le 2$ from the preceding paragraph. In the case
where all the inputs are identical, we know that there is no improvement
possible for $D=1$; in addition, there is no improvement possible for $D=0$,
$D=2=\M-1$ or $D=3=\M$. Thus we find that no improvement is possible with $N\le
3$. Note that this argument does not prove that no improvement is possible with
$\M=3,\ N>3$, because then the result of Sec.~\ref{D1} is not valid.

\section{Numerical testing}
\label{sec:num}
In order to perform numerical testing, we first introduce the
somewhat stronger conjecture
\begin{equation}
\label{eq:simcon}
{\bm{\Gamma}^{(0)}}^\dagger\bm{\Gamma}^{(0)}+{\bm{\Gamma}^{(2)}}^\dagger
\bm{\Gamma}^{(2)}/2-{\bm{\Gamma}^{(1)}}^\dagger\bm{\Gamma}^{(1)} \ge 0.
\end{equation}
As for \eeqref{eq:con}, we implicitly take the inputs to be identical for this
conjecture. If the conjecture given in Eq.~\eqref{eq:con} is \textit{false},
then there exists a vector $\bm{\chi}$ such that
\begin{equation}
    \bm{\chi}^\dagger {\bm{\Gamma}^{(2)}}^\dagger \bm{\Gamma}^{(2)}\bm{\chi}=0,
\end{equation}
but
\begin{equation}
\bm{\chi}^\dagger {\bm{\Gamma}^{(1)}}^\dagger \bm{\Gamma}^{(1)}\bm{\chi}>
\bm{\chi}^\dagger {\bm{\Gamma}^{(0)}}^\dagger\bm{\Gamma}^{(0)} \bm{\chi}.
\end{equation}
In that case,
\begin{equation}
    \bm{\chi}^\dagger{(\bm{\Gamma}^{(0)}}^\dagger\bm{\Gamma}^{(0)}
    +{\bm{\Gamma}^{(2)}}^\dagger\bm{\Gamma}^{(2)}/2-{\bm{\Gamma}^{(1)}}^
    \dagger \bm{\Gamma}^{(1)})\bm{\chi} < 0,
\end{equation}
so the left-hand side of Eq.~\eqref{eq:simcon} has a negative eigenvalue. Thus,
if \eeqref{eq:simcon} is true, then so is \eeqref{eq:con}.

It is more computationally efficient to test Eq.~\eqref{eq:simcon}, because it
does not require a search for vectors $\bm{\chi}$. This expression was tested
for 1000 randomly selected interferometers for values of $N$ from $4$ to $9$,
with interferometer parameters selected according to the Haar measure
\cite{rmt}. Calculations were performed independently for the different values
of $D$. The cases $D=0$, $1$, $N-1$ and $N$ were not tested numerically because
it has been shown analytically that no improvements are possible in those cases.

In no case was a violation of the inequality found within numerical precision.
For $D=2$ the minimum eigenvalues were small positive numbers. For other values
of~$D$ tested the minimum eigenvalues were very small negative numbers on the
order of $-10^{-15}$. Thus the eigenvalues were nonnegative within the precision
of the calculations. This numerical evidence strongly indicates that the
conjecture is true.

\section{Processing of optical qubits}
\label{sec:qu} Now we extend the results to single-rail qubits. A pure SRQ is a
coherent superposition of the vacuum and single-photon states in the same
optical mode: $\ket\phi=\gamma\ket 0+\beta\ket 1$. Similarly to single photons,
SRQs are prone to efficiency losses, so in a practical experimental situation,
there will be an incoherent admixture of the vacuum:
$\hat\rho=E\ket\phi\bra\phi+(1-E)\ket 0 \bra 0$. The state $\hat\rho$ is a
general form of an (inefficient) SRQ; obviously, the inefficient single photon
is a special case of an inefficient SRQ.

We previously investigated the possibilities of modifying the parameters of a
SRQ by means of linear optics and conditional measurements
\cite{berry}. We showed that the appropriate measure of the efficiency of the
SRQ is
\begin{equation} \label{geneff}
\E(\hat\rho) 
 = \frac{|\beta|^2E}{1-|\gamma|^2E}.
\end{equation}
This efficiency can not be increased by linear optical processing on a single
mode. On the other hand, any conversion $(\gamma,\beta,E)\to(\gamma',\beta',
E')$, for which $\E'<\E$, is possible. For inefficient single photons
($\gamma=0,\,\beta=1$), the SRQ efficiency is identical to the single-photon
efficiency $p$.

It is therefore straightforward to generalize our conjecture to SRQs. Consider
the scheme of Fig.~\ref{fig:gen} where the input and output channels carry
qubits of efficiencies $\E_1,\ldots,\E_{\M}$ and $\E_{\rm out}$, respectively.
Condition~(\ref{conjold}) is then equivalent to
\begin{equation} \label{conjqubit}
\sum_{n_1=2}^N \bra{n_1}\hat\rho_{\rm out}\ket{n_1} =0 \implies \E_{\rm out}\le
\E_{\rm max},
\end{equation}
where $\E_{\rm max}$ is the highest of the input generalized efficiencies.

If conjecture \eqref{conjqubit} is true, then conjecture \eqref{conjold} must
also be true as a particular case of the former (the SRQ efficiency cannot be
below the single-photon probability). Conversely, if there were a counterexample
to \eeqref{conjqubit}, there would also be a counterexample to \eeqref{conjold}.
Consider a scheme for processing single-rail qubits with maximum input
efficiency $\E_{\rm max}$ and output efficiency $\E_{\rm out}$. One can use
quantum scissors \cite{scissors1,scissors2} or other schemes (see Appendix) to
produce the SRQ inputs from single-photon sources with efficiency $\E_{\rm
max}+\epsilon$, and produce an inefficient single photon at the output with
efficiency $\E_{\rm out}-\epsilon$ (for arbitrarily small $\epsilon>0$). If the
SRQ processing scheme produced $\E_{\rm out}>\E_{\rm max}$, by selecting
sufficiently small $\epsilon$ one could obtain an improvement in the
single-photon sources. Thus we see that these conjectures are equivalent.

Most of the simplifications derived for processing of single-photon sources also
hold for SRQs. An important exception is that because a SRQ contains coherent
superpositions of different Fock states, we cannot eliminate the possibility of
projections over different photon numbers. Subsequently, it does not make sense
to discuss no-go theorems for particular total photon numbers detected. However,
we can easily see that the no-go theorem for an improvement with $\M\le 2$ still
holds. Indeed, because interconversion between an inefficient photon and a SRQ
is possible with arbitrarily low efficiency loss, an improvement in the case of
$\M\le 2$ input SRQs would imply an improvement for $\M\le 2$ input photons,
which is impossible (Sec.~\ref{M2N3}). However, we can not use this approach to
prove that no improvement is possible with $N=3$, because additional modes are
necessary to transform from single photons to SRQs.

\section{Conclusions}
\label{sec:cnc} We have presented a general form of linear optical processing
for inefficient single photons and single-rail qubits. This processing includes
general measurements, coherent states, and arbitrary numbers of single photons
or single-rail qubits. We have shown that, when searching for schemes that
improve the output efficiency, there are four simplifications that may be made:
\begin{enumerate}
\item The coherent state inputs may be omitted.
\item One may restrict to considering inputs of equal efficiencies.
\item One need only consider projective measurements.
\item For single-photon inputs it is not necessary for the projective
measurement to contain a superposition over different photon numbers.
\end{enumerate}

We have used these simplifications to show that no-go results which hold for
processing of single-photon sources with photon counting also hold when coherent
states and general measurements are allowed.

In addition, we have extended the results to processing of general single-rail
qubits. We have shown, using the single-rail qubit interconversion scheme
\cite{berry}, that the problems of increasing the efficiency of single-photon
sources and the efficiency of single-rail qubits are equivalent. In particular,
we find that it is impossible to increase the efficiency for processing of up to
two single-rail qubits.

It is likely that no increase in the efficiency is possible even for processing
of arbitrary numbers of single-rail qubits. Numerical testing of interferometers
with up to 9~modes found no counterexamples to a somewhat stronger conjecture.
If it is true that no increase in the efficiency is possible, then it would mean
that the efficiency has significant status as a resource for linear optical
processing. This would also be important for linear optical quantum computation,
because it would mean that there is no way of correcting low efficiencies using
linear optics and destructive measurements.

\section*{Acknowledgments} This project has been supported by the Australian
Research Council, iCORE, NSERC, AIF, CIAR, and MITACS.

\appendix
\section*{Appendix A: Example schemes}
\label{sec:ex} 
\begin{figure}
\centerline{\includegraphics[width=5.5cm]{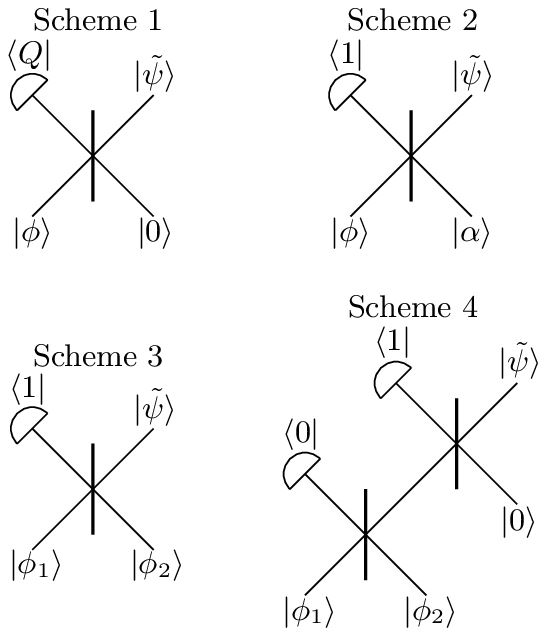}} \caption{\label{figs1}Four
schemes for processing of one or two single-rail qubits. Here
$\ket{\phi_1}=\gamma_1\ket 0+\beta_1\ket 1$, and $\ket{\phi_2}=\gamma_2\ket
0+\beta_2\ket 1$.}
\end{figure}

Here we present a few examples of schemes that can be used to
interconvert between optical qubits of different values and also to obtain a
single-photon state from a single-rail qubit. The four schemes which we consider
are shown in Fig.\ \ref{figs1}. We initially analyze each scheme assuming the
inputs to be pure states, and then state the limits in which the reduction of
the generalized efficiency is minimized. The results for pure single-rail inputs
are summarized in Table 1.

\begin{table*}
\caption{The resulting unnormalized states for the four schemes, as well as the
conditions for a single-photon state. The beam splitter transmission and
reflectivity are denoted by $t$ and $r$, respectively; $t'$ and~$r'$ are the
parameters for the second beam splitter in Scheme~4.} \label{tab1}
\begin{tabular}{|c|c|c|}
    \hline
& $\ket{\tilde\psi}$ & single-photon condition \\
    \hline
Scheme~1 & $(\gamma\braket{Q}{0}+\beta r\braket{Q}{1})\ket{0}+\beta
t\braket{Q}{0}\ket{1}$ &
$\gamma\braket{Q}{0}=-\beta r\braket{Q}{1}$ \\
Scheme~2 & $(\beta r-\alpha\gamma t^*)\ket 0+\beta\alpha(|r|^2-|t|^2)\ket 1$ &
$r/t^*=\gamma\alpha/\beta$ \\
Scheme~3 & $(\beta_1\gamma_2 r-\gamma_1\beta_2 t^*)\ket 0+
\beta_1\beta_2(|r|^2-|t|^2)\ket 1$ & $r/t^*=\gamma_1\beta_2/(\beta_1\gamma_2)$ \\
Scheme~4 & $r'(\beta_1\gamma_2 t+\gamma_1\beta_2r^*)\ket 0
+2r^*t r't'\beta_1\beta_2\ket{1}$ & $t/r^*=-\gamma_1\beta_2/(\beta_1\gamma_2)$ \\
    \hline
\end{tabular}
\end{table*}

\emph{Scheme~1} was proposed and experimentally implemented, for single-photon
inputs, by Babichev et al. \cite{RSP}, and its applications for interconversion
of single-rail qubits were discussed in detail by Berry et al. \cite{berry}. The
input single-rail qubit and the vacuum state entangle themselves at the beam
splitter, generating
\begin{equation}
    \ket{\Psi_{\rm BS}}=\gamma\ket{00}+\beta r\ket{10}+\beta t\ket{01}.
\end{equation}
We then perform a quadrature measurement on mode~1 of $\ket{\Psi_{\rm BS}}$
using a homodyne detector with a certain local oscillator phase. A measurement
result $Q$ is equivalent to projection of mode 1 onto a quadrature eigenstate
$\ket{Q}$, which prepares mode 2 in the single-rail qubit state
\begin{equation}
    \ket{\tilde\psi}=\braket{Q}{\Psi_{\rm BS}}.
\end{equation}
By conditioning on a certain value of $Q$ with a proper local oscillator phase,
one can obtain a qubit of any value, in particular, the single-photon state.
When processing inefficient single-rail qubits, Scheme~1 preserves the SRQ
efficiency in the limit $t\to 1$.

In \emph{Scheme~2}, the vacuum is replaced with a weak coherent field
$\ket\alpha$, and output in mode 2 is conditioned on single photon detection in
mode~1 \cite{catalysis}. The output state has multiphoton components, but these
may be made arbitrarily small by taking the limit of a weak coherent state
$\alpha\ll 1$. In this limit, the initial state approaches
\begin{equation}
    (\gamma\ket 0+\beta \ket 1)\otimes(\ket 0 + \alpha \ket 1).
\end{equation}
Detection of a photon in mode~1 eliminates the two-photon component in the
beam-splitter output, so mode 2 is projected on a single-rail qubit. The limit
of small $\alpha$ implies that the beam splitter reflectivity must also be small
(otherwise the relative fraction of the one-photon component in the output qubit
would vanish). This limit also preserves the generalized efficiency if Scheme~2
is used with inefficient single-rail qubits. However, the probability of success
also approaches zero in this limit.

\emph{Scheme~3} is similar to Scheme~2, except the weak coherent state has been
replaced with another single-rail qubit. The initial state in this case is
$(\gamma_1\ket{0}+\beta_1\ket{1})\otimes(\gamma_2\ket{0}+\beta_2\ket{1})$. There
is a drawback to this scheme, in that the probability for success is zero for
$|\gamma_1|=|\gamma_2|$. \emph{Scheme 4} does not have this problem, but
requires an additional detection \cite{berry1page}.

For Scheme~3, in the limit $\beta_2\to 0$ the output efficiency is
$\E(\hat\rho_1)$. Alternatively, in the limit $\beta_1\to 0$ the output
efficiency is $\E(\hat\rho_2)$. For Scheme~4, if the inputs are identical, in
the limit of low transmissivity for the second beam splitter the output
efficiency is again $\E(\hat\rho)$ ($\hat\rho=\hat\rho_1=\hat\rho_2$). Hence we
find that, for each scheme, the final efficiency is asymptotically equal to the
SRQ efficiency for the input states.

\end{document}